%% file: main.tex
\newcommand{\CLASSINPUTinnersidemargin}{0.75in}
\newcommand{\CLASSINPUToutersidemargin}{0.75in}
\newcommand{\CLASSINPUTtoptextmargin}{0.75in}
\newcommand{\CLASSINPUTbottomtextmargin}{0.75in}

\documentclass[10pt,conference,letterpaper,final]{IEEEtran}

\renewcommand{\IEEEtitletopspaceextra}{0.25in}

\usepackage[T1]{fontenc}
\usepackage[utf8]{inputenc}

\usepackage{algorithmic}
\usepackage{amsmath}
\usepackage{array}
\usepackage[pdftex]{graphicx}
\usepackage{placeins}
\usepackage{todonotes}
\usepackage{url}
\usepackage{xspace}
\usepackage{tikz}
\usetikzlibrary{positioning,calc}

\usepackage{fixltx2e}
\usepackage{dblfloatfix}

\usepackage[noadjust]{cite}

\usepackage{xfrac}
\usepackage{siunitx}
\begin{document}

\input{newcommands}

\title{Robustness from structure: Inference with hierarchical spiking networks on analog neuromorphic hardware}

\author{%
  \IEEEauthorblockN{%
    Mihai~A.~Petrovici\IEEEauthorrefmark{2}\IEEEauthorrefmark{3}\enspace
    Anna~Schroeder\IEEEauthorrefmark{2}\enspace
    \\
    Oliver~Breitwieser\IEEEauthorrefmark{2}\enspace
    Andreas~Gr\"{u}bl\IEEEauthorrefmark{2}\enspace
    Johannes~Schemmel\IEEEauthorrefmark{2}\enspace
    Karlheinz~Meier\IEEEauthorrefmark{2}
  }
  {\footnotesize \tt \{mpedro,\,annasch,\,obreitwi,\,agruebl,\,schemmel,\,meierk\}@kip.uni-heidelberg.de}
  \\
  \vspace{3pt}
  \IEEEauthorblockA{\IEEEauthorrefmark{2} Heidelberg University, Kirchhoff-Institute for Physics, Im Neuenheimer Feld 227, D-69120 Heidelberg}
  \IEEEauthorblockA{\IEEEauthorrefmark{3} University of Bern, Department of Physiology, Bühlplatz 5, CH-3012 Bern}
}

\maketitle

\begin{abstract}
    How spiking networks are able to perform probabilistic inference is an intriguing question, not only for understanding information processing in the brain, but also for transferring these computational principles to neuromorphic silicon circuits.
    A number of computationally powerful spiking network models have been proposed, but most of them have only been tested, under ideal conditions, in software simulations.
    Any implementation in an analog, physical system, be it \emph{in vivo} or \emph{in silico}, will generally lead to distorted dynamics due to the physical properties of the underlying substrate.
    In this paper, we discuss several such distortive effects that are difficult or impossible to remove by classical calibration routines or parameter training.
    We then argue that hierarchical networks of leaky integrate-and-fire neurons can offer the required robustness for physical implementation and demonstrate this with both software simulations and emulation on an accelerated analog neuromorphic device.
\end{abstract}

\input{1_intro}

\input{2_lifbms}

\input{3_spikey}

\input{4_robustness}

\input{5_emulation}

\input{6_conclusions}

\section*{Acknowledgments}

The first two authors contributed equally to this work.
MAP and AS designed and performed the experiments in software and on the Spikey chip.
OB designed the software environment for performing the simulations.
All authors contributed to writing this paper.
We would like to thank Thomas Pfeil and Johannes Bill for technical support with the Spikey chip, as well as Luziwei Leng for providing the model parameters.
We further thank Andreas Baumbach, Johannes Bill, Dominik Dold, Carola Fischer, Vitali Karasenko, Akos Kungl and Walter Senn for their invaluable contribution to the final manuscript.
This research was supported by EU grant \#604102 (Human Brain Project).


\bibliographystyle{IEEEtran}
\input{main.bbl}

\end{document}

%% file: newcommands.tex

\renewcommand{\rm}{{r_\mathrm{m}}\xspace}
\renewcommand{\and}{{\mathrm{and}}\xspace}
\newcommand{\paragraphlb}[1]{\paragraph{#1}\mbox{}\\\vspace{-10pt}}


\newcommand{\bs}[1]{\boldsymbol{#1}}
\newcommand{\minisection}[1]{\paragraph{\emph{#1}}}
\newcommand{\paramtype}[2]{\mathsf{#1_{#2}}}
\newcommand{\sidenote}[1]{\marginnote{\emph{#1}}}
\newcommand{\tb}[1]{\textbf{#1}}
\newcommand{\ti}[1]{\textit{#1}}
\newcommand{\todoin}[1]{\todo[inline]{#1}}


\newcommand{\Bigcap}{\bigcap\limits}
\newcommand{\DKL}[2]{D_\mathrm{KL}\left(#1\parallel#2\right)}
\newcommand{\DKLnorm}[2]{D_\mathrm{KL}^\mathrm{norm}\left(#1\parallel#2\right)}
\newcommand{\Expect}[1]{E\left[#1\right]}
\newcommand{\expect}[1]{\left\langle#1\right\rangle}
\newcommand{\Int}{\int\limits}
\newcommand{\Lim}{\lim\limits}
\newcommand{\Prod}{\prod\limits}
\newcommand{\Sum}{\sum\limits}
\newcommand{\var}[1]{\Var\left[#1\right]}


\newcommand{\ci}[3]{#1 {\perp\!\!\!\perp} #2 \; | \; #3}
\newcommand{\nci}[3]{#1 \centernot{\perp\!\!\!\perp} #2 \; | \; #3}


\newcommand{\abstr}{\mathrm{abstr}\xspace}
\newcommand{\bmErev}{\bm{E}^\mathrm{rev}\xspace}
\newcommand{\bmtausyn}{\bm\tau^\mathrm{syn}\xspace}
\newcommand{\bmueff}{\bm{u}_\mathrm{eff}\xspace}
\newcommand{\boxfct}{\mathrm{box}\xspace}
\newcommand{\Ca}{{\mathrm{Ca}^{++}}\xspace}
\newcommand{\CC}{\rho\xspace}
\newcommand{\Cl}{{\mathrm{Cl}^-}\xspace}
\newcommand{\Cm}{{C_\mathrm{m}}\xspace}
\newcommand{\const}{{\mathrm{const}}\xspace}
\newcommand{\Cref}{{C_\mathrm{ref}}\xspace}
\newcommand{\CVISI}{{\mathrm{CV}_\mathrm{ISI}}\xspace}
\newcommand{\CVrate}{{\mathrm{CV}_\mathrm{rate}}\xspace}
\newcommand{\ddt}{{\frac{d}{dt}}\xspace}
\newcommand{\DeltaT}{{\Delta_\mathrm{T}}\xspace}
\newcommand{\DKLsolo}{{D_\mathrm{KL}}\xspace}
\newcommand{\El}{{E_\mathrm{l}}\xspace}
\newcommand{\Eqn}{\mathrm{Eqn.}\xspace}
\newcommand{\Eqns}{\mathrm{Eqns.}\xspace}
\newcommand{\Er}{{E_\mathrm{r}}\xspace}
\newcommand{\Erev}{E^\mathrm{rev}\xspace}
\newcommand{\Ereve}{{E^\mathrm{rev}_\mathrm{e}}\xspace}
\newcommand{\Erevi}{{E^\mathrm{rev}_\mathrm{i}}\xspace}
\newcommand{\ET}{{E_\mathrm{T}}\xspace}
\newcommand{\gext}{{g_\mathrm{ext}}\xspace}
\newcommand{\gimax}{{g_i^\mathrm{max}}\xspace}
\newcommand{\gl}{{g_\mathrm{l}}\xspace}
\newcommand{\fsyn}{{f^\mathrm{syn}}\xspace}
\newcommand{\gsyn}{g^\mathrm{syn}\xspace}
\newcommand{\gsyne}{{g^\mathrm{syn}_\mathrm{e}}\xspace}
\newcommand{\gsyni}{{g^\mathrm{syn}_\mathrm{i}}\xspace}
\newcommand{\gtot}{{g^\mathrm{tot}}\xspace}
\newcommand{\icb}{\texttt{icb}}
\newcommand{\iext}{{i^\mathrm{ext}}\xspace}
\newcommand{\Iext}{I^\mathrm{ext}\xspace}
\newcommand{\ifmath}{{\mathrm{if}}\xspace}
\newcommand{\Inoise}{I^\mathrm{noise}}
\newcommand{\ipi}{{}^1p_1\xspace}
\newcommand{\irc}{{i^\mathrm{RC}}\xspace}
\newcommand{\Irec}{I^\mathrm{rec}}
\newcommand{\Iref}{{I_\mathrm{ref}}\xspace}
\newcommand{\isyn}{i^\mathrm{syn}\xspace}
\newcommand{\Isyn}{I^\mathrm{syn}\xspace}
\newcommand{\Jsyn}{J^\mathrm{syn}\xspace}
\newcommand{\lambdam}{{\lambda_\mathrm{m}}\xspace}
\newcommand{\Mexc}{M_\mathrm{exc}\xspace}
\newcommand{\Minh}{M_\mathrm{inh}\xspace}
\newcommand{\Na}{{\mathrm{Na}^+}\xspace}
\newcommand{\NBAS}{{N_\mathrm{BAS}}\xspace}
\newcommand{\nne}{{\mathrm{ne}}\xspace}
\newcommand{\NHC}{{N_\mathrm{HC}}\xspace}
\newcommand{\NMC}{{N_\mathrm{MC}}\xspace}
\newcommand{\non}{{\mathrm{\setminus}}\xspace}
\newcommand{\nonk}{{\non k}\xspace}
\newcommand{\NPYR}{{N_\mathrm{PYR}}\xspace}
\newcommand{\nRS}{{n_\mathrm{RS}}\xspace}
\newcommand{\nFS}{{n_\mathrm{FS}}\xspace}
\newcommand{\nusyn}{\nu^\mathrm{syn}\xspace}
\newcommand{\NRSNP}{{N_\mathrm{RSNP}}\xspace}
\newcommand{\otherwise}{\mathrm{otherwise}\xspace}
\newcommand{\pa}{\mathrm{\textbf{pa}}\xspace}
\newcommand{\pflip}{p_\mathrm{flip}\xspace}
\newcommand{\poo}{p_{00}\xspace}
\newcommand{\poi}{p_{01}\xspace}
\newcommand{\pio}{p_{10}\xspace}
\newcommand{\pii}{p_{11}\xspace}
\newcommand{\PSP}{\mathrm{PSP}\xspace}
\newcommand{\pspike}{p_\mathrm{spike}\xspace}
\newcommand{\rl}{{r_\mathrm{l}}\xspace}
\newcommand{\Rtest}{{R^\mathrm{test}}\xspace}
\newcommand{\Rtrain}{{R^\mathrm{train}}\xspace}
\newcommand{\SU}{\tilde I\xspace}
\newcommand{\taubk}{\overline{\tau^\mathrm{b}_k}}
\newcommand{\taudecay}{{\tau_\mathrm{decay}}\xspace}
\newcommand{\taueff}{{\tau_\mathrm{eff}}\xspace}
\newcommand{\taufacil}{{\tau_\mathrm{facil}}\xspace}
\newcommand{\taufall}{{\tau_\mathrm{fall}}\xspace}
\newcommand{\tauinact}{{\tau_\mathrm{inact}}\xspace}
\newcommand{\taum}{{\tau_\mathrm{m}}\xspace}
\newcommand{\tauon}{{\tau_\mathrm{on}}\xspace}
\newcommand{\tauON}{{\tau_\mathrm{ON}}\xspace}
\newcommand{\taurise}{{\tau_\mathrm{rise}}\xspace}
\newcommand{\taurec}{{\tau_\mathrm{rec}}\xspace}
\newcommand{\tauref}{{\tau_\mathrm{ref}}\xspace}
\newcommand{\taustdp}{\tau^\mathrm{STDP}\xspace}
\newcommand{\tausyn}{\tau^\mathrm{syn}\xspace}
\newcommand{\tausyne}{{\tau^\mathrm{syn}_\mathrm{e}}\xspace}
\newcommand{\tausyni}{{\tau^\mathrm{syn}_\mathrm{i}}\xspace}
\newcommand{\tauw}{{\tau_\mathrm{w}}\xspace}
\newcommand{\textmax}{{\mathrm{max}}\xspace}
\newcommand{\textmin}{{\mathrm{min}}\xspace}
\newcommand{\thetaeff}{{\vartheta_\mathrm{eff}}\xspace}
\newcommand{\trise}{{t_\mathrm{rise}}\xspace}
\newcommand{\tfall}{{t_\mathrm{fall}}\xspace}
\newcommand{\tspike}{{t_\mathrm{spike}}\xspace}
\newcommand{\ueff}{{u_\mathrm{eff}}\xspace}
\newcommand{\unif}{{\mathrm{unif}}\xspace}
\newcommand{\ureset}{{u_\mathrm{reset}}\xspace}
\newcommand{\USE}{{U_\mathrm{SE}}\xspace}
\newcommand{\uthr}{{u_\mathrm{thr}}\xspace}
\newcommand{\Vm}{{V_\mathrm{m}}\xspace}
\newcommand{\Vrest}{{V_\mathrm{rest}}\xspace}
\newcommand{\Vspike}{{V_\mathrm{spike}}\xspace}
\newcommand{\Vth}{{V_\mathrm{th}}\xspace}
\newcommand{\Vthresh}{\ET}
\newcommand{\wsyn}{w^\mathrm{syn}\xspace}
\newcommand{\zpi}{{}^2p_1\xspace}


\newcommand{\COBA}{\mathrm{COBA}\xspace}
\newcommand{\Cov}{\mathrm{Cov}\xspace}
\newcommand{\CUBA}{\mathrm{CUBA}\xspace}
\newcommand{\erf}{\mathrm{erf}\xspace}
\newcommand{\for}{\mathrm{for}\xspace}
\newcommand{\sgn}{\mathrm{sgn}\xspace}
\newcommand{\Var}{\mathrm{Var}\xspace}


\newcommand{\aEIFcurrexp}{\mbox{\texttt{aEIF\_curr\_exp}}\xspace}
\newcommand{\IFcondalpha}{\mbox{\texttt{IF\_cond\_alpha}}\xspace}
\newcommand{\NEST}{\mbox{\texttt{NEST}}\xspace}
\newcommand{\Neuron}{\mbox{\texttt{Neuron}}\xspace}
\newcommand{\PyNN}{\mbox{\texttt{PyNN}}\xspace}


\newcommand{\tagarray}{\mbox{}\refstepcounter{equation}$(\theequation)$}
\newenvironment{texttab}[1]
    {\vspace{-10pt}
     \tabulinesep=7pt
     \begin{center}
         \begin{tabu} to 1.013\textwidth {#1}}
        {\end{tabu}
    \end{center}}
    

\def\layersep{2.5cm} 
\def\neuronsep{1.2} 
\tikzstyle{neuron}=[circle,fill=black!25,minimum size=21pt,inner sep=0pt]
\tikzstyle{visible neuron}=[neuron, fill=green!50]
\tikzstyle{hidden neuron}=[neuron, fill=orange!75]

%% file: 1_intro.tex
\section{Introduction}

Over the past decades, research in neural networks has undergone an interesting branching process.
On the one hand, the machine learning community has gradually increased its interest in what were originally brain-inspired neural networks.
These efforts have been crowned by impressive recent success~\cite{lecun2015deep,silver2016mastering}, which has, however been obtained at the price of having strayed away from biologically plausible dynamics.
On the other hand, modern computational neuroscience is pushing for ever more complex and biologically realistic simulations, in the hope to uncover the biological details of information processing in the brain~\cite{markram2015reconstruction}.
Today, these two communities are investigating network models that have little in common with each other.

In the meantime, the neuromorphic community has to master an increasingly difficult balancing act.
At its core, the neuromorphic approach aims to mimic various features of the neocortex in silico. 
For example, an essentially ubiquitous feature of neuromorphic devices is that they are built to emulate spiking neurons~\cite{indiveri2009artificial,pfeil2013six,schemmel2010waferscale,furber2012spinnaker,merolla2014million,benjamin2014neurogrid}.
However, one core argument for building these devices is the hope to use them to unlock the brain's computational power by moving beyond the von Neumann computing paradigms.
Consequently, a driving question for the neuromorphic community might be formulated as follows: is it possible to find relevant applications for spiking neural networks that can then profit from the typical advantages of a physical implementation such as inherent parallelism, high speed and low power consumption?
The findings discussed in this article suggest a promising path towards finding an answer.

This issue is even more pronounced in the case of analog hardware, since it imposes additional constraints that stem from the physics of systems themselves.
As opposed to digital systems, be they von Neumann or neuromorphic, which have the benefit of essentially perfect precision and control, analog systems have to deal with inherent imperfections.
These imperfections concern, on the one hand, the \emph{equations of motion} of the network components, which must obey the physics of the substrate and can therefore only provide an approximation of the target dynamics.
On the other hand, the degree of precision to which the \emph{parameters} of these equations can be tuned certainly depends on the hardware design, but is always fundamentally limited by fixed-pattern variations and temporal noise.
Imperfections in the network dynamics and parameters necessarily distort the behavior of the emulated networks, which usually impairs their performance to some degree~\cite{petrovici2014characterization}.

The question of parameter control (i.e., calibration, post-calibration tuning and training) is an essential one.
It constitutes a perennial challenge for analog neuromorphic system design and operation, and has therefore been often addressed in literature~\cite{indiveri2011neuromorphic,sheik2011systematic,bill2010compensating,petrovici2014characterization}.
A thorough discussion of parameter calibration and in-the-loop training of analog circuits can be found in~\cite{schmitt2016classification}, which represents a complement of the present study.
In the present manuscript, we are mainly concerned with the distortions to the network dynamics that are imposed by the physics of the emulation device and that cannot be directly addressed by, e.g., calibration.

We begin by identifying a Bayesian spiking network model with valuable computational properties (Sec.~\ref{sec:lifbms}).
It is able to learn a probabilistic model of input data and can subsequently be used as both a generative and a discriminative model --- a feature that is difficult to achieve even with abstract neural networks~\cite{leng2016spiking}.
Here, we focus on its discriminative properties.
This model is, in general, susceptible to hardware-induced distortions, as we discuss in detail in Sec.~\ref{sec:spikey}.
As a particular example, we characterize these effects on the Spikey chip~\cite{pfeil2013six} --- the neuromorphic system that we use as an emulation back-end.
Despite our model's ostensible lack of robustness, we argue that, when endowed with a particular, hierarchical connectivity structure, it becomes robust to the studied hardware-induced distortions.
We substantiate this conjecture with both software simulations (Sec.~\ref{sec:robustness}) and hardware emulations (Sec.~\ref{sec:emulation}).
In particular, without any training to compensate for parameter noise on the hardware, we show that the network only loses a relatively small fraction of its initial performance when running on Spikey.
This represents, to our knowledge, the first scalable implementation of a hierarchical probabilistic network in accelerated analog neuromorphic hardware.

%% file: 2_lifbms.tex
\section{LIF-based Boltzmann machines}
\label{sec:lifbms}

\begin{figure*}[t]
    \centering
    \begin{tikzpicture}[
            font=\sffamily,
            line width=1pt,
            scale=1.0,
            ->,
            shorten >=1pt,
            shorten <=1pt,
            >=latex,
            transform shape,
        ]
        \begin{scope}[shift={(-2.8,1.7)},scale=1.0]
            \node[draw, circle, minimum size=2.2cm] (m) at (0,0) {};
            \node[minimum size=0pt] (b) at (0,-1.7) {$\bs{\bar u}^0$};
            \foreach \n in {0,...,2}
            {
                \node[draw, circle, minimum size=0.6cm] (c\n) at (120*\n+90:.6) {};
                \draw (b.north-|c\n) to (c\n.south);
            }
            \path (m) edge [out=30, in=80, looseness=3.5, <->] node[above right] {$\bs w$} (m);
        \end{scope}
        \begin{scope}[shift={(-7.1,-1.2)},font=\small]
            \node[draw,circle,minimum size=.8cm] (top) at (0,0) {$z_2$};
            \node[inner sep=1pt] (b1) [left=.6cm of top] {$b_2$};
            \node[draw,circle,minimum size=.8cm] [below=1.1cm of top] (bottom) {$z_1$};
            \node[inner sep=1pt] (b2) at (b1|-bottom) {$b_1$};
            \draw (b1) to (top);
            \draw[out=220, in=140,looseness=1.2] (top) to (bottom);
            \node[font=\tiny] (mid) at ($(top)!0.5!(bottom)$) {$W_{12}=W_{21}$};
            \draw (b2) to (bottom);
            \draw[out=40, in=-40,looseness=1.2] (bottom) to (top);
        \end{scope}
        \pgfresetboundingbox
        \draw[use as bounding box,inner sep=0pt] node {\includegraphics[width=\textwidth]{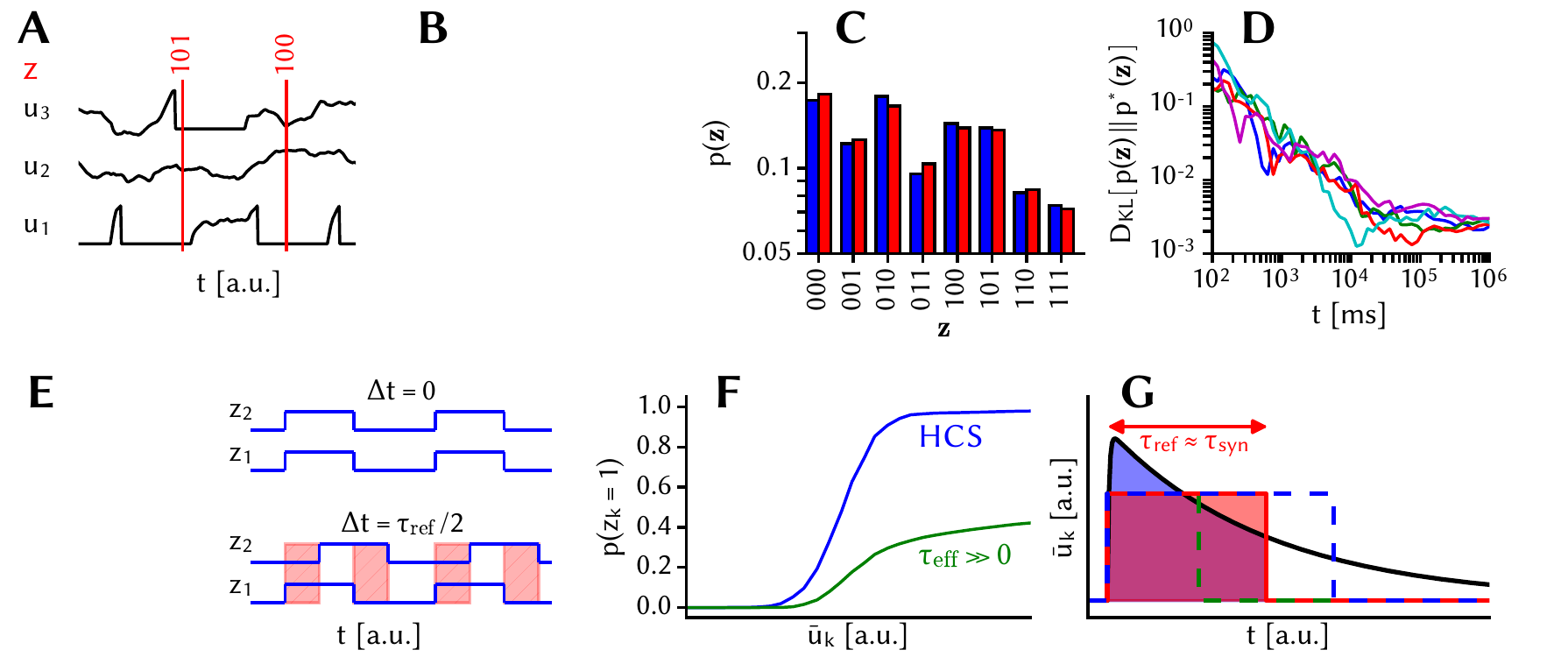}};
    \end{tikzpicture}
    \caption{
        Sampling with LIF neurons and distortions induced by implementation in a physical substrate.
        \tb{(A)} Exemplary membrane potentials from a sampling network of LIF neurons.
                 Each neuron has an associated random variable $z_k \in \{0,1\}$ which is equal to 1 when the neuron is refractory.
        \tb{(B)} Schematic of a recurrent network of LIF neurons in the HCS with a symmetric synaptic weight matrix $\bs w$ and bias potential vector $\bs{\bar u}^0$, which approximates a Boltzmann machine (C, D) with parameters given by Eqns~\ref{eqn:biastrans}~and~\ref{eqn:weighttrans}.
        \tb{(C)} Exemplary state distribution of a 3-neuron network: sampled distribution $p(\bs z)$ (blue) vs.\ target distribution $p^*(\bs z)$ (red) after \SI{e3}{\milli\second}.
        \tb{(D)} Evolution of the Kullback-Leibler divergence $\DKL{p}{p^*}$ over time.
                 Multiple runs with different random seeds are marked with different colors.
        \tb{(E)} Synaptic transmission delays change temporal correlations between the states of different neurons.
                 In this example, we consider two neurons connected with large excitatory weights $w_{12} = -2b_1 = -2b_2$.
                 Without delays ($\Delta t = 0$, top), the network samples correctly from its target distribution $p(0,0) = p(1,1) \approx 0.5$, $p(0,1) = p(1,0) \approx 0$.
                 With relatively large delays ($\Delta t = \tauref / 2$, bottom), the sampled distribution becomes completely different, with $p(0,1) \gg 0$ and $p(1,0) \gg 0$.
                 The wrongly sampled mixed states, marked in red, are a direct consequence of the synaptic transmission delays.
        \tb{(F)} An imperfect high-conductance state ($\taueff \gg 0$) leads to a deviation of the neuronal activation function from its ideal logistic shape.
                 This modifies the sampled distribution by reducing the probability of neurons to spike, especially for positive biases.
        \tb{(G)} Refractory times and synaptic time constants are coupled to ensure that the average interaction between neurons during refractoriness (blue-shaded PSP in the LIF model) has the correct amplitude (red-shaded rectangular PSP in the abstract model), given in Eqn.~\ref{eqn:weighttrans}. 
                 Spike-to-spike variability of refractory times disrupts this equivalence (green and blue dashed lines), effectively modifying the interaction strength $\bs W$.
        }
    \label{fig:1}
\end{figure*}
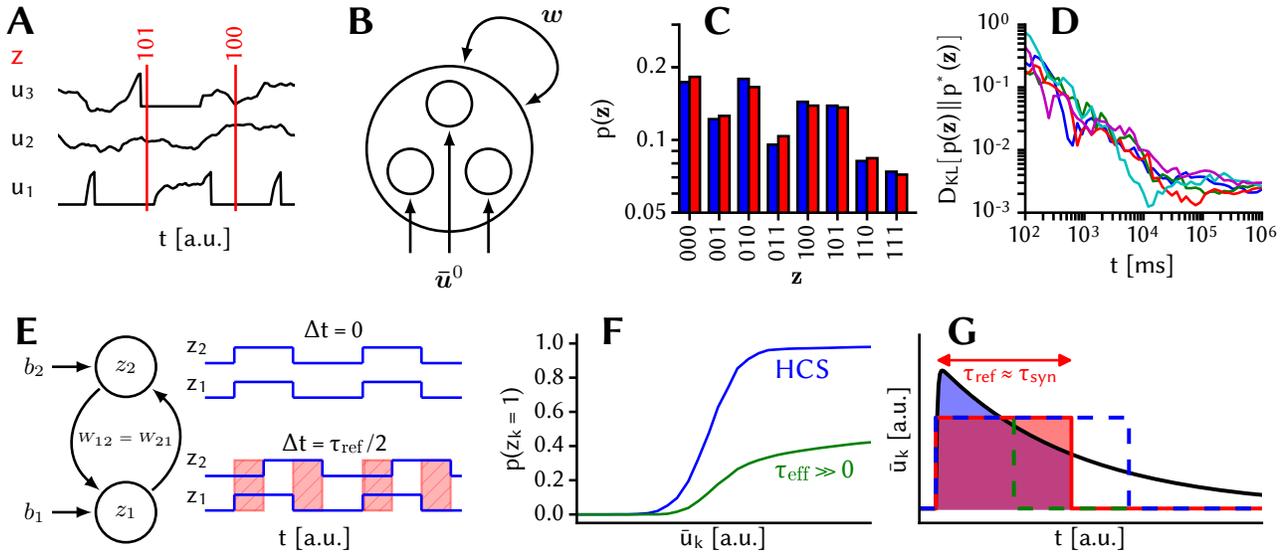

Continued technological advances in large-scale processing (parallel CPU and GPU architectures) have enabled the recent resurgence of artificial neural networks.
Already envisioned for decades as theoretical models of brain-like architectures~\cite{mcculloch1943neuron,rosenblatt1958perceptron}, neural networks now routinely outperform their rival models at pattern recognition tasks~\cite{lecun2015deep}.
Here, we focus on one particular neural network model --- a spiking variant of the Boltzmann machine (BM)~\cite{ackley1985learning}, which has been shown to be compatible with biologically plausible and hardware-implementable spiking neurons~\cite{petrovici2016stochastic}.
We now briefly describe the structure and dynamics of these stochastic networks of leaky integrate-and-fire (LIF) neurons and discuss potential problems that can arise from their implementation in analog hardware.

In the neural sampling framework \cite{buesing2011neural}, a population of $n$ neurons represents a binary random vector $\bs z \in \{ 0,1 \}^n$.
The refractory state of a neuron following a spike at time $t_s$ is chosen to represent the 1-state of the associated random variable (see also Fig.~\ref{fig:1}A):
\begin{equation}
    z_k^{(t)} = \left\{\begin{array}{ll}
                        1 \quad & \quad \text{if $t_s < t < t_s + \tauref$} \quad , \\
                        0 \quad & \quad \text{otherwise} \quad .
                    \end{array}\right.
\end{equation}
In the abstract model of neural sampling, the probability of each neuron to be in the 1-state is given by a logistic activation function
\begin{equation}
    p(z_k=1|\bs z_\nonk) = \sigma (u^\abstr_k) = \frac{1}{1+e^{-u^\abstr_k}} \quad .
    \label{eqn:actfct}
\end{equation}
Such an abstract neuron's membrane potential $u^\abstr_k$ has a resting-state value of $b_k$ and is linearly influenced by the state $\bs z_\nonk$ of all other neurons in the network via (symmetric) synaptic weights $W_{kj}=W_{jk}$:
\begin{equation}
    u^\abstr_k (\bs z_\nonk) = \Sum_{j=1}^n W_{kj} z_j + b_k \quad .
    \label{eqn:uabstract}
\end{equation}
It can then be shown that, under these assumptions, a network of such stochastic neurons will sample from a Boltzmann distribution
\begin{equation}
    p(\bs z) = \frac1Z \exp[-E(\bs{z})] = \frac1Z \exp \left[ \frac12 \bs{z}^T \bs{W} \bs{z} + \bs{z}^T \bs{b} \right] \quad ,
    \label{eqn:jointboltzmann}
\end{equation}
with the partition function $Z$ as a normalization factor. 

In order to achieve similar dynamics with LIF neurons, an equivalent firing regime needs to be established.
In the LIF sampling framework~\cite{petrovici2016stochastic}, each neuron receives two kinds of spiking input: information-encoding input from other neurons in the network and diffuse background input that represents the source of stochasticity, modeled by Poisson sources.
These input spike trains generate two types of current onto the membrane, which we denote by $\Isyn$ and $\Inoise$, respectively:
\begin{equation}
    \Cm \ddt u_k = \gl (\El -u_k) + \Isyn_k + \Inoise_k + \Iext_k \quad .
    \label{eqn:lif}
\end{equation}
Here, $\Cm$ is the membrane capacitance, $\gl$ and $\El$ are the leak conductance and potential, and $\Iext_k$ is an external current that determines the bias $b_k$.
While in general noisy LIF neurons do not have a logistic activation function, as required in Eqn.~\ref{eqn:actfct}, it has been shown that in a high-conductance state the LIF activation function can be well approximated by a logistic function that is scaled with parameters $\alpha$ and $\bar u_k^0$~\cite{petrovici2016stochastic,petrovici2015hcs}:
\begin{equation}
    p(z_k = 1 | \bs z_\nonk) = \sigma \left(\frac{\bar u_k - \bar u_k^0}{\alpha}\right) \quad ,
    \label{eqn:actfctlif}
\end{equation}
where $\bar u_k = \expect{u_k}_t$ represents the noise-free membrane potential of the $k$th neuron.
This equivalence to the abstract model enables an LIF neuron to sample correctly from its conditional distribution $p(z_k|z_\nonk)$.
The translation of the Boltzmann parameters $(\bs W, \bs b)$ in Eqn.~\ref{eqn:jointboltzmann} to the conductance-based LIF domain (synaptic weights $\bs w$, bias potentials $\bs{\bar u}^0$) can then be achieved using the following rules:
\begin{align}
    b_k & = (\bar u^b_k - \bar u^0_k) / \alpha \label{eqn:biastrans}\\
    W_{kj} & = \frac{1}{\alpha \Cm} \frac{w_{kj} \left(\Erev_{kj} - \mu\right)}{\frac{1}{\tausyn} - \frac{1}{\taueff}} \nonumber \\
           & \quad \times \left[\frac{1-e}{e} \!-\! \frac{\taueff}{\tausyn} \left( e^{- \frac{\tausyn}{\taueff}} - 1 \right) \right] ,
    \label{eqn:weighttrans}
\end{align}
where $\mu$ is the mean of the free membrane potential, $\taueff=\Cm/(\gl+\gsyn)$ is the effective membrane time constant in the high-conductance state, $\gsyn$ the total synaptic conductance, $\tausyn$ the synaptic time constant and $\Erev$ the synaptic reversal potential.
This allows a direct mapping of abstract BMs to networks of LIF neurons that sample accurately from their target distribution (Fig.~\ref{fig:1}B-D).

It is important to note that such networks are not merely more complicated replicas of classical machine learning approaches.
In addition to being able to emulate the computational power of traditional Boltzmann machines, these spiking networks can also harness certain biological mechanisms to extend their functionality.
It has, for example, been shown, that when endowed with short-term synaptic plasticity, LIF-based BMs can become good generative models of their learned datasets, while at the same time maintaining a high classification performance when presented with individual data samples~\cite{leng2016spiking}.

Such LIF networks are now amenable to training with any of the established algorithms for BMs.
While backpropagation~\cite{rumelhart1986backprop} is more difficult to implement in a biologically plausible network, other methods exist that are more compatible with Hebbian learning.
The wake-sleep algorithm, in particular, requires each synapse to only have access to the activity of its pre- and its postsynaptic neuron~\cite{hinton2002training}:
\begin{align}
    \Delta w_{ij} &= \eta({\expect{z_i z_j}}_\mathrm{data} - \expect{z_i z_j}_\mathrm{model}) \quad , \\
    \Delta b_i &= \eta(\expect{z_i}_\mathrm{data} - \expect{z_i}_\mathrm{model}) \quad .
\end{align}
This learning rule tries to adapt the activity $\bs z_\mathrm{model}$ of the network in the ``dreaming'' phase, during which it evolves freely, to its activity $\bs z_\mathrm{data}$ in the ``awake'' state, where it is constrained by data, i.e., where some of the units are clamped to particular values.
Despite its simplicity, this learning algorithm can be used to achieve high classification rates on various machine learning datasets~\cite{salakhutdinov2009deep}.

%% file: 3_spikey.tex
\section{Critical distortions in physical implementations}
\label{sec:spikey}

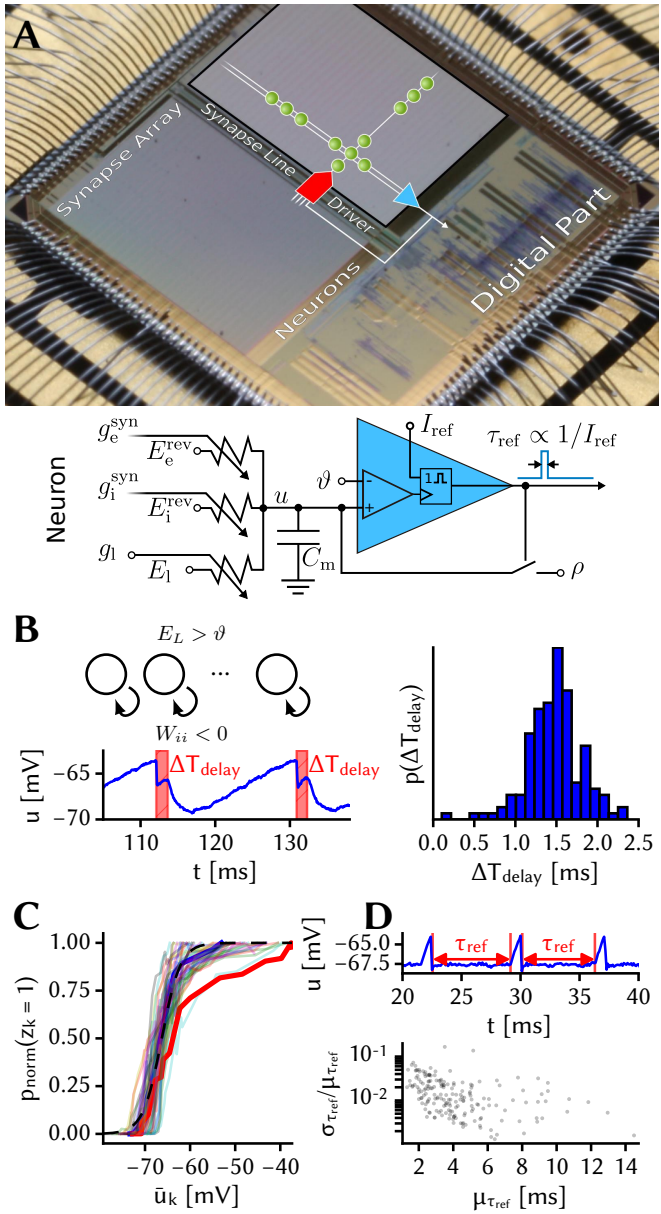
\begin{figure}[t!]
    \centering
    \begin{tikzpicture}[
            font={\sffamily \small},
            scale=1.,
            ->,
            shorten >=2pt,
            shorten <=2pt,
            >=latex,
            transform shape,
        ]
        \pgfdeclarelayer{background layer}
        \pgfsetlayers{background layer,main}
        \draw[use as bounding box,inner sep=0pt, anchor=north west] node {\includegraphics[width=\columnwidth]{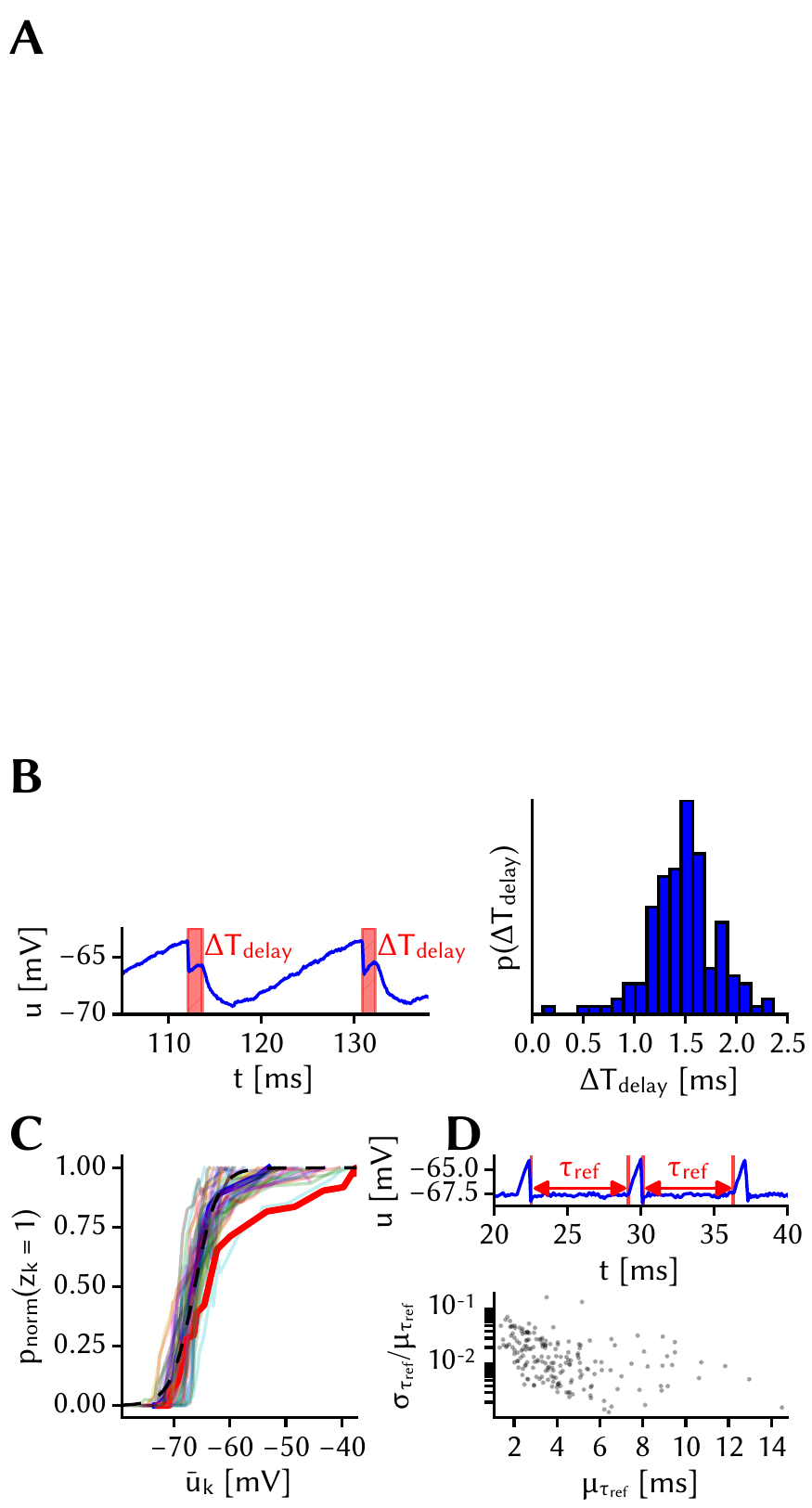}};
        \begin{pgfonlayer}{background layer}
            \newlength\mywidth
            \newlength\myheight
            \pgfextractx{\mywidth}{\pgfpointdiff{\pgfpointanchor{current bounding box}{west}}{\pgfpointanchor{current bounding box}{east}}}
            \pgfextracty{\myheight}{\pgfpointdiff{\pgfpointanchor{current bounding box}{south}}{\pgfpointanchor{current bounding box}{north}}}
            \node[help lines,inner sep=0pt,anchor=north west,rectangle,minimum height=\myheight,minimum width=\mywidth] (spikey) at (current bounding box.north west) {};
            \node[help lines,inner sep=0pt,anchor=north,scale=1.00] (spikey) at (current bounding box.north) {\includegraphics[width=\columnwidth]{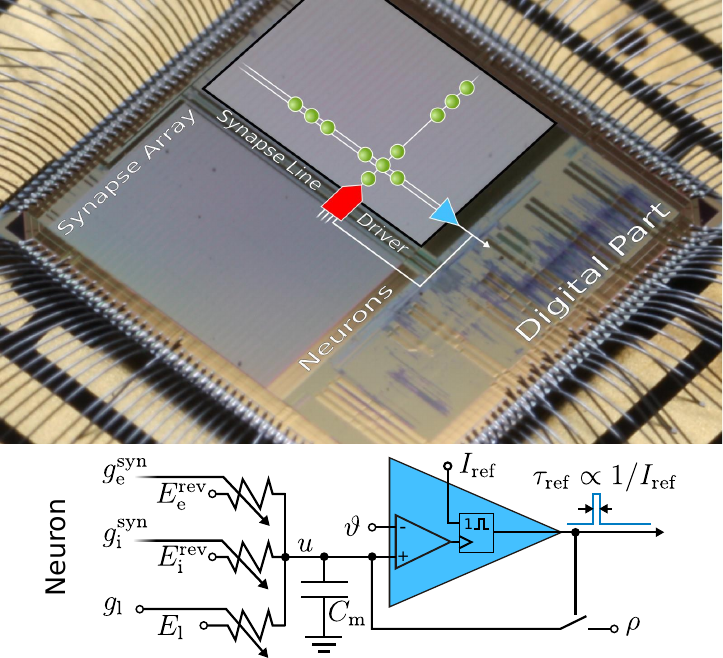}};
        \end{pgfonlayer}
        \begin{scope}[anchor=north west,shift={(spikey.south west)},shift={(1.15,-0.8)},line width=1pt, node distance=.25cm, scale=0.87]
            \tikzstyle{bendy}=[out=-30, in=-70, looseness=6]
            \tikzstyle{neuron}=[draw, circle, minimum size=.6cm]
            \node[neuron] (c1) {};
            \path (c1) edge [bendy] (c1);
            \node[neuron, right=of c1] (c2) {};
            \path (c2) edge [bendy] (c2);
            \node[right=of c2, font=\large] (dots) {...};
            \node[neuron, right=of dots] (c3) {};
            \path (c3) edge [bendy] (c3);
            \node (top) [above=of $(c1)!0.5!(c3)$,yshift=.1cm] {$E_L > \vartheta$};
            \node (bottom) [below=of $(c1)!0.5!(c3)$,yshift=-.4cm] {$W_{ii} < 0$};
        \end{scope}
    \end{tikzpicture}
    \caption{
        Characterization measurements for the employed neuromorphic system.
        \tb{(A)} \emph{Top:} Neuromorphic Spikey chip with overlaid sketch of neural network components (taken from~\cite{pfeil2013six}).
                 \emph{Bottom:} Simplified schematic of a single neuron.
        \tb{(B)} \emph{Top left:} Synaptic delays were measured by recurrently connecting each neuron to itself through an inhibitory synapse.
                 \emph{Bottom left:} The relatively sharp onset of the inhibitory PSP allows a precise measurement despite the temporal noise on the membrane potential.
                 \emph{Right:} Synaptic delay distribution of 192 neurons for a single synapse driver.
        \tb{(C)} Activation functions of 44 Spikey neurons (thin solid lines) compared to the nearly ideal logistic activation function achieved in the high-conductance state (dashed line).
                 Two exemplary activation functions are drawn with thicker lines: the blue and red activation functions belong to neurons with short and long $\taueff$, respectively. 
        \tb{(D)} \emph{Top:} Refractory times were measured by choosing a suprathreshold leak potential for all neurons and subtracting the reset-to-threshold first passage time from the interspike interval.
                 \emph{Bottom:} Relative spike-to-spike variability of the refractory time vs. mean refractory time for 192 Spikey neurons.
        }
        \label{fig:2}
\end{figure}

The distribution that an LIF network samples from is uniquely determined by the neuro-synaptic dynamics and parameters.
Any deviation from the model specification will alter the sampled distribution and, in general, restrict the network's ability to perform correct inference in the learned sample space.
Mapping this model to an imperfect physical substrate is therefore not straightforward.
In this article, we study three types of distortions of network dynamics that are caused by mapping to an analog silicon substrate.

First, we consider spike transmission delays.
Since we are using point neurons, we can describe all delays as being synaptic delays.
Many analog neuromorphic devices, including the one we use later on, are mixed-signal systems, meaning that spikes are transmitted digitally.
Consequently, digitization, transport of the digital data, and the conversion back to the analog domain in the synapses contribute to synaptic delays.
While these delays may be short in terms of wall-clock time, they become particularly critical in accelerated systems.
In such systems, the neuronal and synaptic dynamics that define the characteristic time scale on which the network evolves can be orders of magnitude smaller than in biology, potentially entering the range of synaptic transmission delays~\cite{petrovici2014characterization}.
Regardless of the exact nature of a network performing neural sampling, in order for each neuron to be able to calculate its correct conditional distribution $p(z_k|\bs z_\nonk)$, the information gathered by a neuron from its incoming PSPs must coincide with the true state of the corresponding presynaptic neurons, as required by Eqn.~\ref{eqn:uabstract}.
This temporal coincidence is disrupted by delays, which thereby distort the sampled distribution, as exemplified in Fig.~\ref{fig:1}E.

Second, while most neuromorphic systems have controllable neuron and synapse parameters, these can only be configured within a certain range and resolution.
As an example, consider the membrane time constant $\taum$ of a neuron.
This time constant can be considered to define the reaction speed of a neuron to external stimuli.
In neuromorphic systems, $\taum$ is usually configured with an adjustable leak conductance, which can not become arbitrarily large.
Such a limit in the reaction speed of neurons can impair the functionality of the entire network.
For LIF neurons, a large $\taum$ slows the saturation of the activation function (Fig.~\ref{fig:1}F) and thereby distorts the logistic shape (Eqn.~\ref{eqn:actfctlif}) required for sampling from Boltzmann distributions.

Third, temporal noise can also affect computation.
Depending on the particular in-silico implementation, any analog system will be subject to some degree of temporal noise on all of its electronic signals, including those that directly influence neuro-synaptic dynamics.
In our particular case, the largest temporal noise component affects refractory times $\tauref$.
Since the relevant neuron and synaptic circuits are physically disconnected on the chip, the spike-to-spike variation of $\tauref$ is independent from the synaptic time constant, which can be considered as fixed.
Consequently, the state $z_k(t)$ of a neuron will not coincide anymore with the information it transmits via PSPs to its postsynaptic partners (Fig.~\ref{fig:1}G), leading to a distortion of the sampled probability distribution in a conceptually similar manner as synaptic delays do.

In this study, we use the Spikey single-chip system as a physical emulation substrate~\cite{pfeil2013six}.
This mixed-signal device combines analog components for modeling membrane and synapse dynamics with digital circuitry for the spike-based communication.
Fig.~\ref{fig:2} shows a photo of the device, along with a sketch of the neuron circuit which illustrates the origin of the three distortive effects discussed above.

The overlay in Fig.~\ref{fig:2}A shows how a spike emitted by a neuron (blue triangle) travels through communication buses (white line) to a synapse driver (red pentagon), which generates a voltage ramp that is fed into the synapse array (green circles).
Inside the synapse, the voltage ramp is converted to an approximately exponential signal that is added to the total synaptic conductance of the neuron circuit.
This sequence of processing stages causes the effective synaptic delays seen in Fig.~\ref{fig:2}B.

The neuron schematic in Fig.~\ref{fig:2}A explains the cause of non-logistic activation functions and noisy refractory times.

All reversal potentials are connected to the membrane by conductances that saturate at a certain amplitude.
The maximum total conductance defines a minimum achievable effective membrane time constant, which limits the gain of the LIF activation function, as seen in Fig.~\ref{fig:2}C.

The duration of the refractory time is determined by a monoflop controlled by a current $\Iref \propto 1/\tauref$.
In order to offset, at least to some extent, the effect of delays (see Fig.~\ref{fig:1}E) we require long refractory times, i.e., small currents, which cause some of the transistors in the monoflop to leave saturation and operate in a sub-threshold regime.
This transition is accompanied by an increase in the relative amplitude of temporal noise, which increases the variability of $\tauref$.
This represents the primary cause of the large spike-to-spike fluctuations of the refractory time seen in Fig.~\ref{fig:2}D.

Having identified the origins for critical distortions in physical implementations of LIF sampling, we next turn to the central question of this study: is it possible to recover the computational capabilities of LIF sampling networks by finding a network architecture that is robust to these substrate-induced effects?

%% file: 4_robustness.tex
\section{Robust hierarchical LIF networks}
\label{sec:robustness}

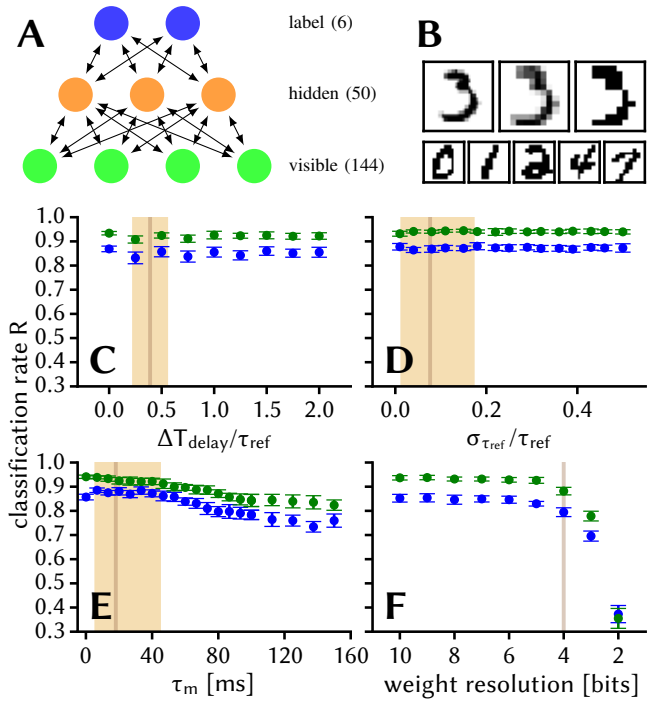
\begin{figure}[t!]
    \centering
    \begin{tikzpicture}[
            transform shape
        ]
        \newlength\halfplot
        \draw[use as bounding box,inner sep=0pt] node {\includegraphics[width=\columnwidth]{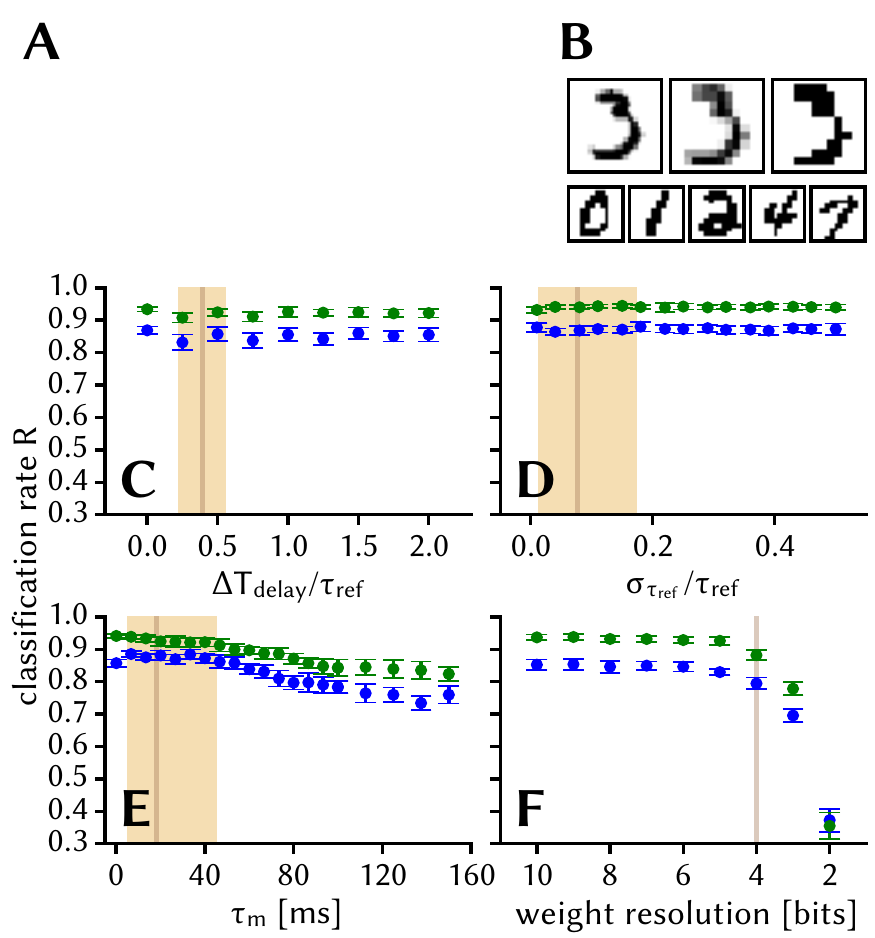}};
        \begin{scope}[shift={(-3.8cm,2.475cm)},
            font=\scriptsize,
            ->,
            shorten >=2pt,
            shorten <=2pt,
            >=latex
            ]
            \def\layersep{.95cm} 
            \def\neuronsep{.95} 
            \tikzstyle{neuron}=[circle,minimum size=13pt,inner sep=0pt]
            \tikzstyle{visible neuron}=[neuron, fill=green!75]
            \tikzstyle{hidden neuron}=[neuron, fill=orange!75]
            \tikzstyle{label neuron}=[neuron, fill=blue!75]
            \def\numlab{2} 
            \def\numhid{3} 
            \def\numvis{4} 
            \pgfmathsetmacro{\max}{max(\numvis, \numhid, \numlab)}
            \pgfmathsetmacro{\texthoroffset}{.1pt}

            \foreach \x in {1,...,\numvis}
            \pgfmathparse{((\max - \numvis)*0.5 + \x - 1)*\neuronsep}
            \node[visible neuron] (V\x) at (\pgfmathresult,0) {};

            \foreach \x in {1,...,\numhid}
            {
                \pgfmathparse{((\max - \numhid)*0.5 + \x - 1)*\neuronsep}
                \node[hidden neuron] (H\x) at (\pgfmathresult, \layersep) {};
            }

            \foreach \x in {1,...,\numlab}
            {
                \pgfmathparse{((\max - \numlab)*0.5 + \x - 1)*\neuronsep}
                \node[label neuron] (L\x) at (\pgfmathresult, \layersep*2) {};
            }

            \foreach \source in {1,...,\numvis}
                \foreach \dest in {1,...,\numhid}
                    \draw[<->] (V\source) -- (H\dest);

            \foreach \source in {1,...,\numhid}
                \foreach \dest in {1,...,\numlab}
                    \draw[<->] (H\source) -- (L\dest);

            \node[anchor=west,right=\texthoroffset of V\numvis] (vl) {visible (144)};
            \node[anchor=west] at (vl.west|-H\numhid) {hidden (50)};
            \node[anchor=west] at (vl.west|-L\numlab)  {label (6)};
        \end{scope}
    \end{tikzpicture}
    \caption{
        Hierarchical LIF networks and their robustness to hardware-induced distortions.
        \tb{(A)} Structure of the studied LIF network.
        \tb{(B)} \emph{Top row:} Exemplary training sample obtained from the MNIST dataset after resolution reduction and binarization.
                 \emph{Bottom row:} Exemplary training samples from all other classes.
        \tb{(C)--(F)} Influence of simulated hardware-induced distortions on the classification performance of the network from A.
                     Error bars represent the standard deviation over multiple runs with different random seeds.
                     Green: performance on training data.
                     Blue: performance on test data.
                     Brown: mean value and standard deviation of the respective parameter measured on Spikey (see Fig.~\ref{fig:2}).
        }
    \label{fig:3}
\end{figure}

The general framework of LIF sampling does not impose any restrictions on the network topology apart from the requirement of a zero-diagonal symmetric synaptic weight matrix $W_{ii}=0, \; \bs W = \bs W^T$.
However, imposing further restrictions on connectivity is of practical use.

When building a network that is able to learn and generalize from data, a rather natural hierarchization consists in subdividing the network into a layer representing the ``visible'' data, one or more hidden layers that recognize common features of data samples, and a final classification layer that assigns each sample a particular category or label.
Indeed, this has been the guiding principle behind hierarchical neural networks, from multilayer perceptrons to deep convolutional nets~\cite{haykin2004comprehensive}.
In the case of BMs, the further removal of lateral connections within a layer has proven particularly beneficial for learning~\cite{hinton2006reducing}.
The resulting networks are so-called restricted Boltzmann machines (RBMs) and can be emulated by LIF networks with appropriate parameters as described in Sec.~\ref{sec:lifbms}.

The core insight of our present work is that hierarchical LIF networks that emulate RBMs exhibit notable robustness to the hardware-induced distortions discussed above.
In this section, we argue why this is the case and demonstrate this robustness with software simulations of such a network with 3 layers (Fig.~\ref{fig:3}A).

Synaptic delays and noisy refractory times have similar effects on the sampled distribution.
However, the nature of the information flow in LIF-based RBMs is expected to counter them both simultaneously.
When presented with unambiguous input data, the mean firing rates of the visible neurons $v_k$ are fixed; in our application, for example, they encode the grayscale values $g_k \in \{0, 1, \dots, 255\}$ of pixels in the input image:
$p(v_k=1) = g_k/255$.
In this regime, spike transmission delays have no effect, as the visible layer essentially operates in a rate-based mode for which time shifts do not matter.
In this operating mode, the refractory noise is also averaged out.

Transmission delays and noisy refractory times remain critical for the interaction between hidden and label neurons. 
However, this interaction is comparatively weak in our 3-layer architecture (see Fig.~\ref{fig:3}A).
In real-world scenarios, the label space typically has a much smaller dimensionality than the input space.
Each hidden neuron therefore receives input from many visible neurons but only from few label neurons.
Therefore, even though the visible-to-hidden synaptic weights are approximately as large as those between the hidden and label layer, the summed input from the visible layer is completely dominant by virtue of sheer numbers.
Therefore, as the hidden layer is mostly driven by the input layer, the distorted interaction between the hidden and label neurons is likely to become insignificant.

The finite membrane time constant, on the other hand, can affect neurons in all layers and can not be neglected.
However, this effect can be countered, at least to some extent, by the nature of the sampled distribution in well-trained networks.
Wake-sleep training has the effect of carving of deep troughs in the network's energy landscape $E(\bs z)$.
These energy minima (probability maxima) correspond to particular patterns in each of the network layers, which are local attractors in the state space.
Thus, if the deviations in the sampled distribution are small, the attractor landscape will not change significantly.
Consequently, when the visible layer is clamped to input data, the above layers are still likely to fall into the corresponding attractor state, thus conserving the classification performance.

We tested these predictions in a series of software simulations.
We trained a 3-layer LIF-based RBM (Fig.~\ref{fig:3}A) on a reduced version of the MNIST dataset~\cite{lecun1989recognition}.
To ensure compatibility with the Spikey chip, the network size was restricted to 144 visible, 50 hidden and 6 label neurons.
The $12 \times 12$ pixel images belonging to 6 digit classes (``0'', ``1'', ``2'', ``3'', ``4'', ``7'') were produced by first reducing the MNIST digit resolution, followed by binarization of the pixel values (Fig.~\ref{fig:3}B).
For each class, both the training and the test set consisted of 20 randomly chosen images.
The training consisted only of layer-by-layer pre-training with a wake-sleep-style algorithm~\cite{salakhutdinov2010learning}.
We have deliberately refrained from fine-tuning the weights with, e.g., backpropagation, in order to maintain compatibility with Hebbian plasticity.

Fig.~\ref{fig:3}C-E show the simulated effects of the three hardware-induced distortion mechanisms discussed above.
As expected, neither synaptic transmission delays (Fig.~\ref{fig:3}C) nor variability of refractory times (Fig.~\ref{fig:3}D) affected the performance of the network significantly.
Over a surprisingly large range of membrane time constants, the classification rate remained almost unaffected.
Only after the activation function became significantly distorted by large $\taum$ did the attractor landscape change significantly enough to cause a decay in the classification rate (Fig.~\ref{fig:3}E).
Overall, within the parameter ranges of the Spikey chip, our network remained only weakly affected by the studied mechanisms.

However, since in a later step the network was mapped to the hardware without any further training, we needed to also consider the effect of discretized synaptic weights.
By default, synaptic weights on the Spikey chip are only controllable up to 4-bit precision~\cite{pfeil2013six}.
It is important to note that this does not pose a fundamental problem to networks of this type; the effects of weight discretization can be countered by appropriate in-the-loop training, as discussed in, e.g.,~\cite{esser2016convolutional,schmitt2016classification}.
Here, we only take this effect into account as a preparation of the hardware experiments in Sec.~\ref{sec:emulation}.
Fig.~\ref{fig:3}F shows the effect of weight discretization on the network's classification performance.
For the Spikey chip, the performance decay lies at approximately \SI{5.6}{\percent}.
Note that this effect is significantly larger than the effects caused by each of the other distortion mechanisms.

A combined simulation of all distortive effects was used to provide a reference for the later emulation on Spikey.
All effects were simulated with amplitudes corresponding to values measured on Spikey (blue bars in Fig.~\ref{fig:3}C-F, see also Fig.~\ref{fig:2}).
In the ideal, undistorted case, the LIF network had a classification performance of
\begin{equation}
    \begin{aligned}
        \Rtrain &= \SI{93.4 +- 0.9}{\percent} \\
        \Rtest &= \SI{86.6 +- 1.7}{\percent} \quad ,
        \label{eqn:swideal}
    \end{aligned}
\end{equation}
which was reduced to
\begin{equation}
    \begin{aligned}
        \Rtrain &= \SI{90.7 +- 1.7}{\percent} \\
        \Rtest &= \SI{78.1 +- 1.5}{\percent}
        \label{eqn:swspikey}
    \end{aligned}
\end{equation}

when all distortive effects were simultaneously present.
A comparison to Fig.~\ref{fig:3}F shows that most of this performance decay was due to the 4-bit weight discretization.

%% file: 5_emulation.tex
\section{Neuromorphic implementation}
\label{sec:emulation}

\begin{figure}[t!]
    \centering
    \includegraphics[width=\columnwidth]{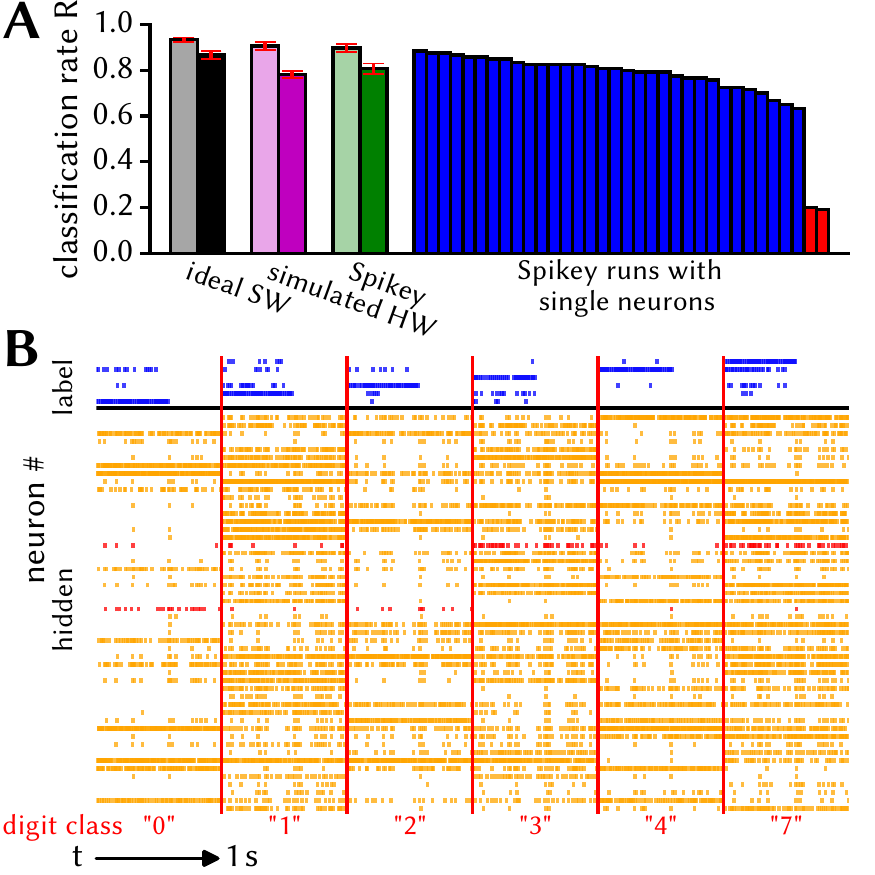}
    \caption{
        Study of a direct-to-hardware mapping of the hierarchical LIF network from Fig.~\ref{fig:3}A.
        \tb{(A)} Classification performance.
                 Black: Software simulation of a distortion-free LIF network (cf. leftmost data points in Fig.~\ref{fig:3}C-E).
                 Purple: Software simulation of the LIF network with all distortion mechanisms being present simultaneously, with amplitudes and variances as measured on Spikey (cf. areas in Fig.~\ref{fig:3}C-E highlighted in brown).
                 Green: Hardware emulation of the hidden layer with software evaluation of the label layer.
                 Blue/red: Emulation of the hidden layer with repeated use of single neurons, followed by software evaluation of the label layer.
                 The two ``bad'' neurons marked in red were not well configurable and therefore performed at chance level.
        \tb{(B)} Exemplary spike trains of a subset of neurons in the LIF network with the hidden layer running on hardware.
                 Spike trains belonging to the two ``bad'' neurons from A are marked in red.
        }
    \label{fig:4}
\end{figure}

The mapping of the network to Spikey required a series of modifications, which we discuss in the following.

In the previous section, we argued that visible neurons essentially operate in a rate-based mode during clamping.
This allows the activity $\nu_i$ of each visible neuron to be modeled as an effective bias
\begin{equation}
    \tilde b_{ki} = p(v_i=1) \cdot w_{ki} \quad ,
\end{equation}
to the $k$th hidden neuron, where $v_i$ represents the state of the $i$th visible neuron and $w_{ki}$ the synaptic weight between the $i$th visible and the $k$th hidden neuron.
The complete visible layer can then be omitted altogether and replaced by an effective bias $\tilde b_k$ for each hidden neuron:
\begin{equation}
    \tilde b_k = \Sum_i \tilde b_{ki} + b_k \quad ,
\end{equation}
where $b_k$ is the original bias of the $k$th hidden neuron.
In our particular case, since we use $12\times12$--pixel binarized images and have set all biases to zero during training in order to simplify the transition to hardware, this reduces to
\begin{equation}
    \tilde b_k = \Sum_{i=1}^{144} w_{ki} v_i \quad .
\end{equation}

On the chip, biases are implemented as high-frequency regular spike trains connected to the hidden neurons with weights $w^b_k$.
For an arbitrary synaptic kernel scaled with $w^b_k$, the average effect of a regular spike train on the membrane potential of an LIF neuron is proportional to $w^b_k$.
Therefore, $\tilde b_k$ can be controlled, within the imposed 4-bit precision, by appropriately configuring $w^b_k$.
Note that these spike trains also need to be routed across the chip (Fig.~\ref{fig:2}), so this does not circumvent synaptic delays.

For the hidden layer, we have chosen those 50 neurons on the chip which responded best to the bias stimulus described above.
Only half of the chip was used for these experiments in order to simplify on-chip routing.

The label layer was implemented in software.
Spikes produced by the hidden layer were fed into six label neurons simulated with NEST \cite{gewaltig2007nest}.
With this, we essentially broke the hidden$\rightarrow$label$\rightarrow$hidden feedback loop, but as we argued in Sec.~\ref{sec:robustness}, it should not significantly affect the classification performance of the network.
Furthermore, this allowed a more detailed investigation of the quality of single neurons on the chip, as discussed below.
The label assigned by the network to the input image was determined by the label neuron which produced the most spikes during the clamping period.

Fig.~\ref{fig:4} shows the classification performance of this setup, along with the spike trains from several exemplary classification runs.
The performance of the hardware implementation was
\begin{equation}
    \begin{aligned}
        \Rtrain &= \SI{89.8 +- 1.8}{\percent} \\
        \Rtest &= \SI{80.7 +- 2.3}{\percent} \quad .
        \label{eqn:hwspikey}
    \end{aligned}
\end{equation}
Within the error margins, this corresponds very well to the reference software simulations (\ref{eqn:swspikey}).
The slightly better average classification can be attributed to the explicit selection of the 50 hidden neurons.
Indeed, this result not only confirms the robustness of our network model, but also highlights its robustness to various other hardware-induced distortions that we did not explicitly account for, such as parameter noise and crosstalk \cite{pfeil2013six}.
Furthermore, this implementation is surprisingly robust even towards few neurons having strongly deviant firing characteristics, as discussed below.

In our network model, hidden neurons are not laterally interconnected.
Furthermore, as the label layer was simulated in software, there was also no label-mediated lateral interaction between hidden neurons.
Therefore, it was possible to emulate the entire network with one Spikey neuron at a time.
In an alternative emulation setup, a sequence of $k=(1,\dots,50)$ emulation runs containing a single hardware neuron was performed.
In the $k$th run, the hardware neuron was configured to represent the $k$th hidden neuron by receiving the corresponding input spike train.
The output spike trains from these runs were aggregated and fed into the label layer, as before.
This experiment was repeated for a subset of 38 out of the 50 selected hardware neurons, with the results plotted as thin bars in Fig.~\ref{fig:4}A.

The overall performance of each neuron quantifies its quality for the task at hand.
The main reason for the differences between the neurons is the shape of their activation function, some of which can be seen in Fig.~\ref{fig:2}C.
Some neurons perform poorly because their activation function is too shallow, thus strongly skewing the sampled distribution.
At the other extreme, a very steep activation function is also detrimental, because the resolution of the synaptic weights does not permit an arbitrarily fine-grained tuning of effective weights and biases.
Note, in particular, how two of the neurons perform at chance level (Fig.~\ref{fig:4}A, red bars).
However, the existence of such neurons does not appear to have a strong effect on the classification performance of the network as a whole (Fig.~\ref{fig:4}A, purple vs. green bar).

%% file: 6_conclusions.tex
\section{Discussion}

One of the most important challenges for analog neuromorphic computing is the design of neural network architectures that are robust to hardware-induced distortions of network dynamics and parameters.
In this paper, we have argued that hierarchical spiking sampling networks emulating restricted Boltzmann machines are inherently resistant to such distortions.
We have studied three specific distortion mechanisms that are, in general, strongly disruptive to the ongoing computation in sampling LIF networks: synaptic transmission delays, variability of refractory times and saturating membrane conductances.
Despite their apparent sensitivity, we have shown how a hierarchical topology shapes the information flow in a way that makes them largely resilient to these effects.
The results obtained in software simulations were also confirmed in experiments on an accelerated analog neuromorphic device.
Furthermore, in addition to being robust to the studied distortion mechanisms, our network model also displayed an encouraging degree of resilience to other hardware-induced effects which can not be quantified as systematically as the studied ones.

The choice of our neuron model (LIF) was made, on one hand, for analytical tractability, but, more importantly, due to the fact that this model represents a common denominator for many other spiking neuron models (Hodgkin-Huxley, Izhikevich, AdEx).
With appropriate parameter choices, all of these models can achieve dynamics that are close to those required for sampling.
Furthermore, the LIF model represents a de-facto standard in neuromorphic engineering~\cite{indiveri2009artificial,pfeil2013six,schemmel2010waferscale,furber2012spinnaker,merolla2014million,benjamin2014neurogrid}.

Altogether, the observed properties of these networks encourage further theoretical and experimental investigation.
Here, we only studied relative performance losses, so we only used a relatively small network and a very small dataset.
Software simulations show that larger-scale versions of these networks enable efficient and powerful inference in more complex data spaces \cite{leng2016spiking}.
It will be interesting to see whether such large networks remain as robust to hardware-induced distortions as their smaller instantiations studied here.
Large-scale accelerated analog devices are already in place \cite{schemmel2010waferscale} and will be able to accomodate these experiments.
With our proposed architecture and rather conservative clamping schedule of 1 biological second per image (Fig.~\ref{fig:4}B), the currently achieved acceleration factor of about \num{e4} will, for example, enable the classification of the full MNIST dataset within \SI{1}{\second} of wall-clock time.

Although we only used our hierarchical LIF networks for classification, they can also be used as generative models to perform, for example, pattern completion.
The nature of the energy landscape in these networks suggests that their generative properties could also be robust to hardware-induced distortions.
Since a clear image corresponds to a deep mode in the energy landscape, small distortions in the sampled distribution are unlikely to strongly disrupt the generative properties of the network.
Probabilistic switching between different modes when ambiguous input is present can then be facilitated by short-term plasticity as shown in \cite{leng2016spiking}, a mechanism that is readily available on several accelerated neuromorphic platforms \cite{pfeil2013six,schemmel2010waferscale}.

In this paper, we have deliberately refrained from further training of the hardware-emulated networks.
However, it is expected that training the hardware ``in the loop'' will significantly improve classification.
The idea behind in-the-loop training is to iteratively alternate between a forward pass on hardware, during which the emulated network activity is recorded, and a backward pass in software, where the network parameters are updated by, e.g., error backpropagation.
Recent studies have demonstrated, both in software \cite{lillicrap2016random} and on analog neuromorphic hardware \cite{schmitt2016classification}, that the parameter updates need not be precise, but only approximately follow the gradient of the likelihood function.
Furthermore, accelerated systems currently in development \cite{friedmann2016demonstrating,schemmel2016accelerated} also implement powerful on-chip learning solutions.
Such architectures will not only enable accelerated classification, but, even more importantly, accelerated learning of the network parameters.